\documentclass[aps,prl,twocolumn,groupedaddress,nofootinbib]{revtex4}
\usepackage{graphicx}
\usepackage{amssymb}
\usepackage{color}
\usepackage{soul}

\begin{document}

\title{Effect of dispersion interactions on the properties of LiF in condensed phases}
\author{Dario Corradini$^{1,2}$, Dario Marrocchelli$^3$, Paul A. Madden$^4$ and Mathieu Salanne$^{1,2}$}
\address{$^1$ Sorbonne Universit\'es, UPMC Univ Paris 06, UMR 8234, PHENIX, F-75005, Paris, France}
\address{$^2$ CNRS, UMR 8234, PHENIX, F-75005, Paris, France}
\address{$^3$ Department of Materials Science and Engineering, Massachusetts Institute of Technology, Cambridge, MA, USA}
\address{$^4$ Department of Materials, University of Oxford, Parks Road, Oxford OX1 3PH, UK}
\begin{abstract}
Classical molecular dynamics simulations are performed on LiF in the framework of the polarizable ion model. The overlap-repulsion and polarization terms of the interaction potential are derived on a purely non empirical, first-principles basis. For the dispersion, three cases are considered: a first one in which the dispersion parameters are set to zero and two others in which they are included, with different parameterizations. Various thermodynamic, structural and dynamic properties are calculated for the solid and liquid phases.
The melting temperature is also obtained by direct coexistence simulations of the liquid and solid phases.
Dispersion interactions appear to have an important effect on the density of both phases and on the melting point, although the liquid properties are not affected when simulations are performed in the $NVT$ ensemble at the experimental density.
\end{abstract}
\maketitle
\section{Introduction}
Dispersion interactions between two chemical entities arise from the instantaneous correlation of the fluctuations of their electron densities~\cite{london1937a}. They are the only attractive interactions in noble gases and are therefore at the origin of the existence of their pure condensed phases~\cite{vanderwaals1910a}. But dispersion interactions are generally weaker than the other intermolecular interactions in molecular systems and their role in systems for which stronger attractive forces are present is hard to establish. 

In density functional theory (DFT) calculations, dispersion interactions are difficult to capture. Several methods have been developed, in which they are treated directly by devising new functionals~\cite{vydrov2010a,vydrov2010b} or effectively by adding an analytical term to existing functionals~\cite{grimme2004a}. Following these developments, it was shown that dispersion interactions impact structure and dynamics in a variety of systems; for example in water at ambient~\cite{lin2009b} or supercritical conditions~\cite{jonchiere2011a}. But the most affected quantity is usually the density, which is understimated by as much as 20~\% in water when dispersion interactions are not accounted for~\cite{schmidt2009a}. Similar deviations are found in crystalline systems~\cite{allen2011a}. In ionic materials, the effect of dispersion is expected to be smaller due to the presence of a strong attractive Coulombic interaction between species with different charges. This leads to strong charge ordering effects, and the overlap-repulsion interaction sets the distance of closest approach between two ions. The polarization (induction) interaction also plays a well-identified role and it is at the origin of the stabilization of non-trivial crystal structures and of the cross-linking of the coordination shells of multivalent cations in ionic melts~\cite{salanne2011c}. Nevertheless, it was shown by Kirchner {\it et al.} that in ``room-temperature'' ionic liquids, the inclusion of dispersion interactions in DFT calculations modifies the vibrational and dynamic properties of the system~\cite{grimme2012a,malberg2012a,pensado2012a}. 

DFT-based molecular dynamics is however limited to short simulation times and small system sizes, so that it is hard to establish clearly the role of dispersion on a wide range of properties. Classical molecular dynamics is an alternative option, provided that the force field accurately separates the various contributions from Coulombic, repulsion, polarization and dispersion interactions. In recent years, we have developed a series of methods which make possible the derivation of such force fields for inorganic ionic materials~\cite{aguado2003b,madden2006a,rotenberg2010a,salanne2012b}. In short, we use a polarizable ion model, and all the parameters except the ones concerning dispersion are fitted to standard \emph{condensed phase} DFT calculations. The emphasis on ``condensed phase" here is because the physical properties of ions are strongly affected by their coordination environment and the use of {\it ab initio} data on isolated ions or on the interaction energies of small clusters leads to hopelessly inadequate force-fields. This applies particularly to the dispersion coefficients (and polarizabilities). For our procedure the reference data includes both the dipole moments on each ion and the total force which is exerted on them, allowing the polarization and the repulsion terms to be fitted independently (formal charges are used for the ions, which automatically sets the Coulombic interaction). In addition, some parameters such as the condensed phase polarizability of the ions can directly be calculated~\cite{heaton2006b,salanne2008e,molina2011a}. The dispersion coefficients have to be treated separately, because on one hand the dispersion effects are not yet reliably represented in the DFT calculations and on the other they contribute relatively weakly to the forces on the ions, which means that they do not strongly affect the quality of the fit to the DFT data with the classical force-field.

In the present work, we perform simulations of LiF in the crystal and liquid phases. We compare the thermodynamic, structural and dynamic properties obtained with two different sets of condensed phase dispersion coefficients, obtained either from the {\em Coupled Hartree-Fock} (CHF) theory~\cite{fowler1985b} or from a recent method~\cite{rotenberg2010a,silvestrelli2008a} involving the use of the {\em Maximally Localized Wannier Functions} (MLWFs)~\cite{marzari1997a,marzari2012a}  in DFT, as well as with the case where the dispersion interactions are set to zero. LiF was chosen because it is a material for which we expect that the effect of the dispersion interactions is among the lowest. Dispersion is negligible for the Li$^+$ cation due to its small radius. Also, F$^-$ is the halide anion with the smallest polarizability.  

\section{Numerical methods}

\subsection{Polarizable Ion Model}
The polarizable ion model includes Coulombic, dispersion, overlap repulsion and polarization components~\cite{madden1996a}.
First the  Coulombic
term is:
\begin{eqnarray}
V^{\rm Coul}=\sum_{i<j}\frac{q^i q^j}{r^{ij}}
\end{eqnarray}

\noindent where $q^i$ is the charge on ion $i$, and formal charges are  used
throughout. The dispersion component includes dipole-dipole and dipole-quadrupole
terms
\begin{eqnarray}
V^{\rm disp}=-\sum_{i<j}\left(f^{ij}_6(r^{ij}) \frac{C_6^{ij}}{(r^{ij})^6}+f^{ij}_8(r^{ij}) \frac{C_8^{ij}}{(r^{ij})^8} \right)
\end{eqnarray}
\noindent where $C_6^{ij}$ ($C_8^{ij}$) is the dipole-dipole  (dipole-quadrupole)
dispersion coefficient, and $f^{ij}_n$ are damping
functions \cite{tang1984a}, describing the short-range penetration correction to
the asymptotic multipole expansion of dispersion \cite{stone-book} ($f^{ij}_n(0)=0$ and
$f^{ij}_n(\infty)=1$). They take the form
\begin{eqnarray}
f^{ij}_n(r^{ij}) = 1-{\rm e}^{-b_n^{ij} r^{ij}}\sum_{k=0}^n \frac{(b_n^{ij} r^{ij})^k}{k!}
\end{eqnarray}

\noindent and the parameters $b_n^{ij}$ represent the distance at which the
correction begins to be taken into account. The repulsion overlap component is given by
\begin{eqnarray}
V^{\rm rep}=\sum_{i<j} B^{ij}{\rm e}^{-a^{ij}r^{ij}}
\end{eqnarray}

Finally the polarization part of the potential includes charge-dipole and dipole-dipole terms:
\begin{eqnarray}
V^{\rm pol}&=&\sum_{i<j}\left(q^i\mu_\alpha^j g_D^{ij}(r^{ij})-q^j\mu_\alpha^i g_D^{ji}(r^{ij})\right)  {\mathbb T}_\alpha^{(1)} \nonumber\\
       & & -\sum_{i<j}\mu_\alpha^i\mu_\beta^j {\mathbb T}_{\alpha\beta}^{(2)}+\sum_i \frac{1}{2\alpha^i}\mid \vec{\mu}^i\mid^2 \, .
\end{eqnarray}

\noindent Here ${\mathbb T}_\alpha^{(1)}$ and ${\mathbb T}_{\alpha\beta}^{(2)}$ are
the charge-dipole and dipole-dipole interaction tensors and $\alpha^i$ is the
polarizability of ion $i$. Again, we include some short-range effects
which are due to the high compression of the ions in condensed ionic materials~\cite{fowler1985a,domene2002a,jemmer1999a}.  These short-range induction effects are
straightforwardly included through the use of damping functions similar to the ones
used in the dispersion term:
\begin{eqnarray}
g_D^{ij}(r^{ij}) = 1-c_D^{ij}{\rm e}^{-b_D^{ij} r^{ij}}\sum_{k=0}^4 \frac{(b_D^{ij} r^{ij})^k}{k!}.
\end{eqnarray}

Here $c_D^{ij}$ is a parameter that reflects the
amplitude of this damping at ion $j$ due to the presence of $i$ and $b_D^{ij}$
again is a range parameter.

The instantaneous values of the dipole moments $\{\vec{\mu}^i\}_N$ are  obtained
by minimization of $V^{\rm pol}$ with respect to these variables: they will therefore
depend on the instantaneous positions of neighboring ions and consequently change
at each timestep in an MD run. The interaction potential can therefore be seen to
contain three additional degrees of freedom (induced dipoles), which describe the
state of the electron charge density of the ions. When calculating the forces on
the ions in an MD simulation, these electronic degrees of freedom should have their
``Born-Oppenheimer'' values, which minimize the total potential energy, for every
atomic configuration. We search for the ground state configurations of these
degrees of freedom at each time step, using a conjugate gradient routine
\cite{numericalrecipes}.
The dynamics is thus similar to the so-called Born-Oppenheimer first-principles
molecular dynamics, as implemented, for example, in the CP2K code \cite{cp2k}.

We perform an Ewald summation of all electrostatic interactions and also of
dispersion \cite{chen1997a}. Thus, all those interactions are free from truncation
errors. The short-range repulsion, which is an exponentially decaying function of
distance, is however truncated beyond a distance equal to, at maximum, half the shortest dimension of the simulation cell.

\subsection{Parameterization}

\begin{table}
\begin{center}
\begin{tabular}{|l | c | c| c| c |  }
\hline
Ion pair & $A^{ij}$ & $a^{ij}$ & $b^{ij}_D=b^{ji}_D$ & c$^{ij}_D$    \\
\hline
F$^-$-F$^-$ & 282.3 & 2.444 & -- & --  \\
F$^-$-Li$^+$ & 18.8 & 1.947 & 1.834 & 1.335  \\
Li$^{+}$-Li$^{+}$& 1.0 & 5.0 & -- & --  \\
\hline
\end{tabular}
\end{center}
\caption{\label{tab:param-repulsion} Fitted parameters for the repulsion and polarization terms for LiF (atomic units). The fluoride polarizability was set to 7.9~atomic units, and the Li$^+$ is not polarizable.}
\end{table}

\begin{table}
\begin{center}
\begin{tabular}{|l | c | c|  }
\hline
Setup & $C_6$ & $C_8$ \\
\hline
No dispersion & 0.0 & 0.0 \\
CHF & 15.0 & 150.0 \\
Wannier & 26.3 & 87.7 \\
\hline
\end{tabular}
\end{center}
\caption{\label{tab:param-dispersion} F$^-$-F$^-$ dispersion parameters for the three simulation setups (atomic units). $b_6^{ij}=b_8^{ij}=1.9$, and the other $C_6$ and $C_8$  coefficients (for Li$^+$-Li$^+$ and Li$^+$-F$^-$) are set to zero in all cases.}
\end{table}
  
Parameters for the repulsion and polarization terms were obtained from a generalized force-fitting procedure described elsewhere~\cite{aguado2003b,salanne2012b}. This approach has been used successfully for a series of oxide and fluoride materials~\cite{burbano2011a,marrocchelli2012a,salanne2012c,tangney2002a,han2010a}. Note that the PBE functional was used to provide the reference data to which the parameters were fitted~\cite{perdew1996a}. From our experience, this functional can be considered as ``dispersion-free'' in the case of ionic materials. The obtained parameters are provided in Table \ref{tab:param-repulsion}. As for the dispersion term, three different cases are tested. In the first case considered, we set it to zero. In the second one (``CHF''), we have taken $C_6$ and $C_8$  parameters which reproduce the Coupled Hartree-Fock calculation by Fowler {\it et al.}~\cite{fowler1985b}. Finally, in the third case (``Wannier''), these parameters were determined from a calculation of the MLWFs~\cite{marzari1997a,marzari2012a} using the procedure described in references~\cite{rotenberg2010a,salanne2012b}. The three sets of dispersion coefficients are provided in Table \ref{tab:param-dispersion}. 

\begin{figure}
\begin{center}
\includegraphics[width=\columnwidth]{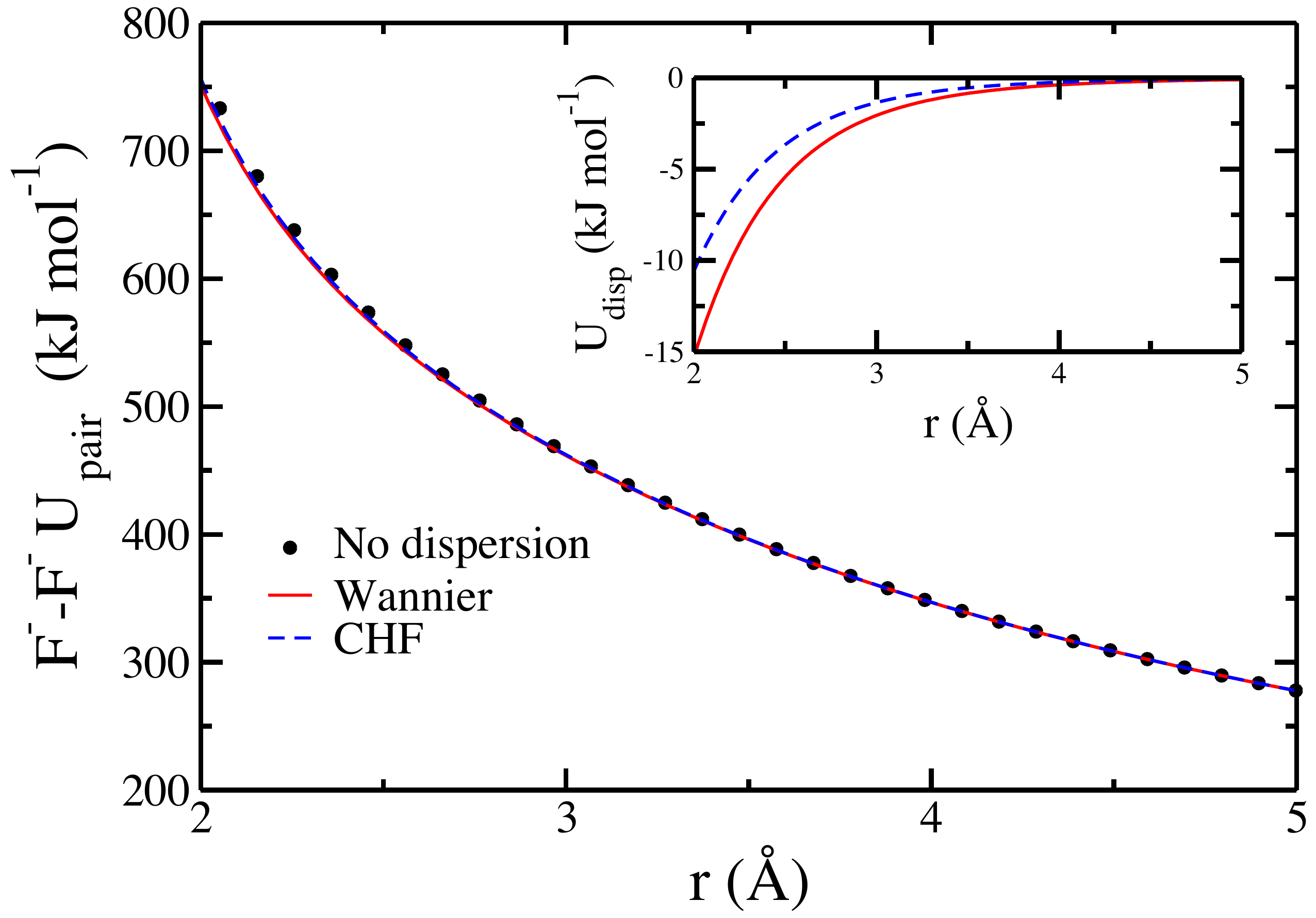}
\end{center}
\caption{\label{fig:potentiel} Sum of the three pairwise additive terms for the F$^-$-F$^-$ interaction (inset: dispersion term only). The distance range corresponds to the extent of the first peak of the F-F partial radial distribution function in liquid LiF (shown in Figure \ref{fig:rdf}).}
\end{figure}

The only interaction which differs between the different cases is the F$^-$-F$^-$ one. The sum of the three pairwise additive terms is plotted for each parameters set in Figure \ref{fig:potentiel}, in the region corresponding to first neighbour typical distances. The inset shows the dispersion term only. It appears obvious that the dispersion term only provides a tiny contribution (between 10 and 15~kJ mol$^{-1}$ for the shortest distance) to the total interaction potential (which takes values ranging between 450 and 750 kJ mol$^{-1}$ in the same region).  

\subsection{Simulation details}

We consider the rocksalt (B1) $8 \times 8 \times 8$ (2048 LiF units) crystal structure for the solid phase. We build the crystal using the LiF lattice parameter at ambient conditions, $a=4.026$ \AA. The solid thus obtained is equilibrated in the $NPT$ ensemble~\cite{martyna1994a}, at $T=300$~K and ambient pressure, using a thermostat time constant $\tau_T=1$~ps and an isotropic barostat time constant $\tau_P=2$~ps. 
Equilibrium is rapidly reached over 10 ps simulations. We then simulate the crystal 
for accumulating the values of the density over another 20~ps. 

For the liquid phase, we do two series of simulations. On the one hand we perform $NVT$ runs at $T=1200$~K and at the experimental density, at ambient pressure, $\rho_{\rm exp}=1.77$~g~cm$^{-3}$~\cite{janz1968a}. The simulation cell for the liquid phase contains 500 LiF units. The temperature is controlled by a Nos\'e-Hoover thermostat chain~\cite{nose1984a,hoover1985a} with a time--constant $\tau_T=10$~ps.  

On the other hand we also perform $NPT$ simulations at $T=1200$~K and at ambient pressure in order to determine the equilibrium density produced by each of the three sets of parameters. 
In this case we use 432 LiF units. The pressure is controlled by applying a barostat~\cite{martyna1994a}, with $\tau_P$ and $\tau_T$ set to 10 ps.

The melting point calculations are performed by direct simulations of the coexisting 
crystalline and molten phase in the $NPT$ ensemble
~\cite{espinosa2013a,conde2013a,hong2013a}. The simulation cell contains 896 LiF
units and it has a large aspect ratio, initially $L_x=L_y=L_z/4$. First we perform $NVE$
simulations that allow us to roughly estimate the melting temperature as described in 
Ref.~\cite{lanning2004b}. Starting from the final ionic configurations from the $NVE$ runs, which contain a crystalline and liquid region close to coexistence, we then proceed to perform $NPT$ direct coexistence simulations at 
ambient pressure and at several temperatures in proximity of the melting temperature 
estimated by the $NVE$ method. The thermostat time constant is set to 1 ps and the 
anisotropic barostat time constant is set to 2 ps. The angles between the cell vectors 
are fixed to $\pi/2$, so that the symmetry of the cell remains orthogonal, but the lengths of the sides of the simulation cell are able to fluctuate independently in order that the calculated pressure tensor is, on average, isotropic with the diagonal elements equal to ambient pressure. The time step 
for the integration of the equations of motion is 1 fs. As the simulation progresses, we monitor the extent of the crystalline and liquid regions over a 500~ps run. For temperature significantly above or below $T_m$, the whole cell tends towards becoming molten or crystalline and this enables us to identify a temperature at which it is possible to conduct a run at which the extent of the liquid and solid regions can coexist over the length of the simulation.

\section{Results}

\subsection{Density}


\begin{table}
\begin{center}
\begin{tabular}{|l | c | c|  }
\hline
Setup & $\rho^{\rm sol}$ (g cm$^{-3}$)& $\rho^{\rm liq}$ (g cm$^{-3}$) \\
\hline
Experiment & 2.64 & 1.77 \\
No dispersion & 2.32 & 1.50  \\
CHF & 2.38 & 1.62  \\
Wannier& 2.42 & 1.69  \\
\hline
\end{tabular}
\end{center}
\caption{\label{tab:densities} Equilibrium densities of solid LiF at $T$~=~300~K and of liquid LiF at $T$~=~1200~K.}
\end{table}

 Equilibrium densities extracted from the $NPT$ simulations of solid and liquid LiF at respective temperatures of 300 and 1200~K are provided in Table \ref{tab:densities}. The comparison with experiments shows that similar differences are obtained in both cases when dispersion effects are included. The largest difference is observed when they are omitted, with an underestimation of 15.3~\% for $\rho^{\rm liq}$. Such a deviation is of the same order of magnitude as that obtained in liquid water at room-temperature from DFT calculations using PBE or BLYP functionals~\cite{schmidt2009a}. When the dispersion effects are included, the liquid density remains underestimated, by 4.5~\% and 8.5~\% in the Wannier and CHF cases, respectively. We note that in order to reproduce more accurately the experimental density, larger values would be necessary for $C_6$ and $C_8$ parameters. Such an underestimation of the density was already observed in our previous work on LiF-ThF$_4$, while a perfect agreement was obtained for LiF-NaF-KF and NaF-ZrF$_4$ mixtures~\cite{salanne2009a,dewan2013a} (both studies used the CHF parameters). This shows that although the polarizable ion model transferability for multiple physical and chemical conditions is well established, a completely transferable model would require more complex functional form, in which ``environmental effects'' are taken into account. These environmental effects have mainly been studied for the polarizability~\cite{jemmer1998a,wilson1999a}, and since dispersion effects follow a similar dependence on the electronic cloud extension as does the former, equivalent functional forms could be employed. 

\subsection{Liquid properties at fixed density}
\begin{figure}
\begin{center}
\includegraphics[width=\columnwidth]{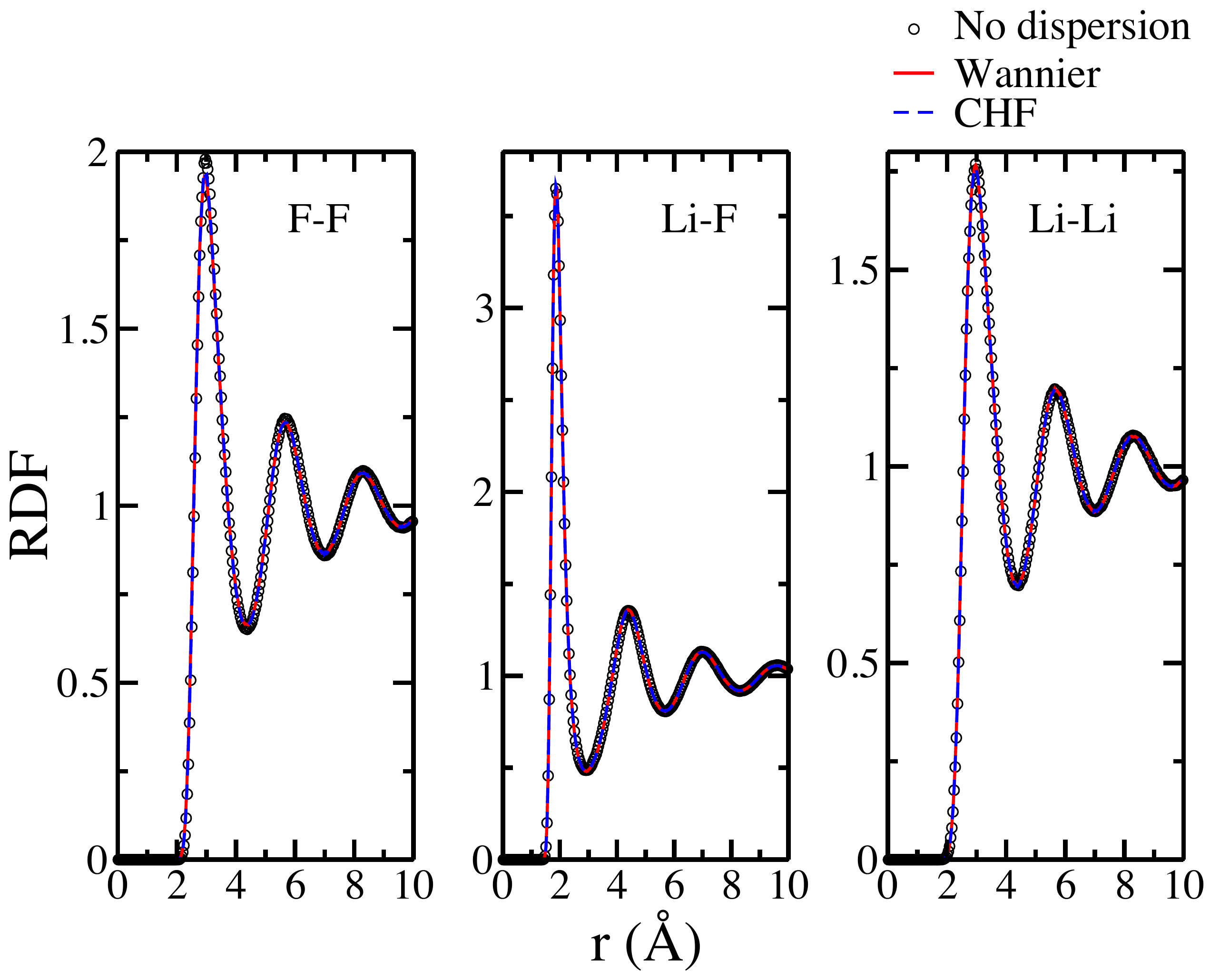}
\end{center}
\caption{\label{fig:rdf} Radial distribution functions in liquid LiF. Simulations were performed at 1200~K, at the experimental density $\rho_{\rm exp}=1.77$~g cm$^{-3}$.}
\end{figure}

In liquid water, the dispersion interaction was shown to impact strongly the structure and dynamics at ambient~\cite{lin2009b} and supercritical~\cite{jonchiere2011a} conditions, even at fixed density. In particular, the first peaks of the O-O and O-H radial distribution functions and the diffusion coefficients were shown to vary substantially. This is not the case for LiF: as can be seen in Figure \ref{fig:rdf}, no difference is observed for any of the three partial radial distribution functions. This is because in such a simple molten salt, the structure mostly arises from a competition between the Coulombic interaction, which induces strong charge ordering, and the overlap-repulsion which sets the first-neighbour distance. The dispersion term is much weaker than those two and therefore does not change the structure. 

\begin{figure*}[ht!]
\begin{center}
\includegraphics[width=0.9\textwidth]{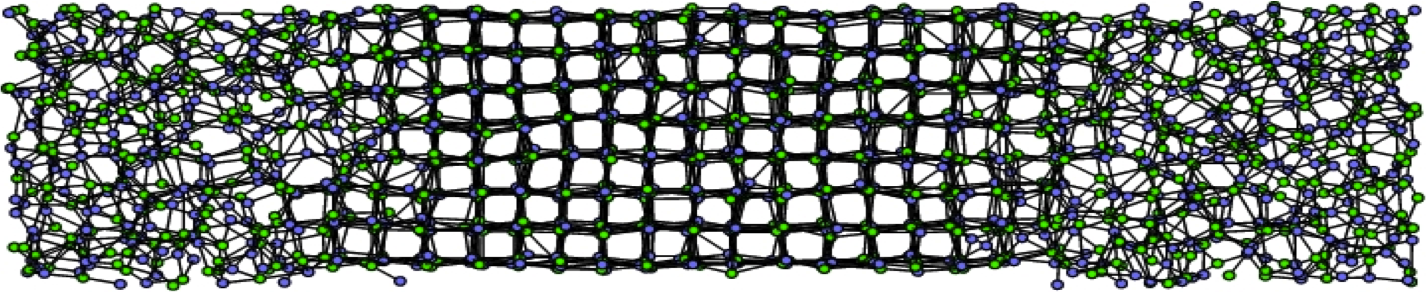}
\end{center}
\caption{\label{fig:interface} Illustration of the simulation cell used to determine the LiF melting point (green: F$^-$, blue: Li$^+$).}
\end{figure*}

The diffusion coefficients are extracted from molecular dynamics simulations using Einstein's relation,
 i.e.  from the long-time slope of  mean
 squared displacement:
 \begin{equation}
 D\left(\alpha\right)=\lim_{t\rightarrow\infty}\frac{1}{6N_{\alpha}t}\sum_{i\in\alpha,i=1}^{N_{\alpha}}\left\langle \left|{\bf r}_{i}(t)-{\bf r}_{i}\left(0\right)\right|^{2}\right\rangle ,
 \end{equation}
 where $N_{\alpha}$ is the total number of atoms of type $\alpha$,
 and ${\bf r}_{i}\left(t\right)$ is the position of atom $i$ of type $\alpha$
 at time $t$. We obtain similar values for the three simulation setups, i.e. $7.4 \cdot 10^{-5}$ and $10.0 \cdot 10^{-5}$ cm$^2$ s$^{-1}$ for F$^-$ and Li$^+$ ions, respectively. Note that these values are in excellent agreement with the experimental ones~\cite{saroukanian2009a,levesque2013b}. Again, this situation is very different from that observed in liquid water, in which a diffusion coefficient two to three times larger was obtained when dispersion effects were included in DFT-based molecular dynamics~\cite{lin2009b} (although these numbers might be mitigated due to the use of short simulation times). In conclusion, due to the predominance of Coulombic and repulsion interactions, in contrast with the case of water, dispersion interaction does not play a major role on the physico-chemical properties of molten LiF at a given density.

\subsection{Solid-liquid interface}

The melting point is usually considered as a very stringent test of a force-field, since it depends on the relative free energies of the crystal and liquid phases. The melting point can be determined by creating a cell containing the solid and
liquid at coexistence. The interface has to be constructed
by combining two separate bulk simulations of the crystalline
and molten phases each equilibrated at the same
pressure and the estimated melting temperature~\cite{lanning2004b,aguado2005a}. The cell parameters for the liquid phase have to match those of the
crystalline simulation.  The two cells are then placed in
a supercell, with a large aspect ratio, as shown on Figure \ref{fig:interface}. A first estimation of the melting point is obtained {\it via} short simulations performed in the $NVE$ ensemble, before longer simulations are performed in the $NPT$ ensemble. In the first stage,
the system is run for a short period of time (1 ps)
in a constant volume simulation with the thermostat set
at the estimated coexistence temperature (or by regular
velocity rescaling) to remove excess energy caused by
bringing the solid and liquid together. Great care must
be taken during these initial stages to ensure that one or
other phase is not destroyed. Once the excessive
relaxation energy has been removed, the system can be
run on for a further 50 ps in a $NVE$ simulation
to ensure complete equilibration. We then perform a 50~ps production run, again in the $NVE$ ensemble, during which we determine the average temperature and pressure. Other ($T$,$P$) coexistence points can then be sampled by performing additional simulations where the initial kinetic energy of the system is rescaled to a value above that of coexistence and then the system is allowed to re-equilibrate in an $NVE$ simulation~\cite{lanning2004b}.   

\begin{figure}[ht!]
\begin{center}
\includegraphics[width=\columnwidth]{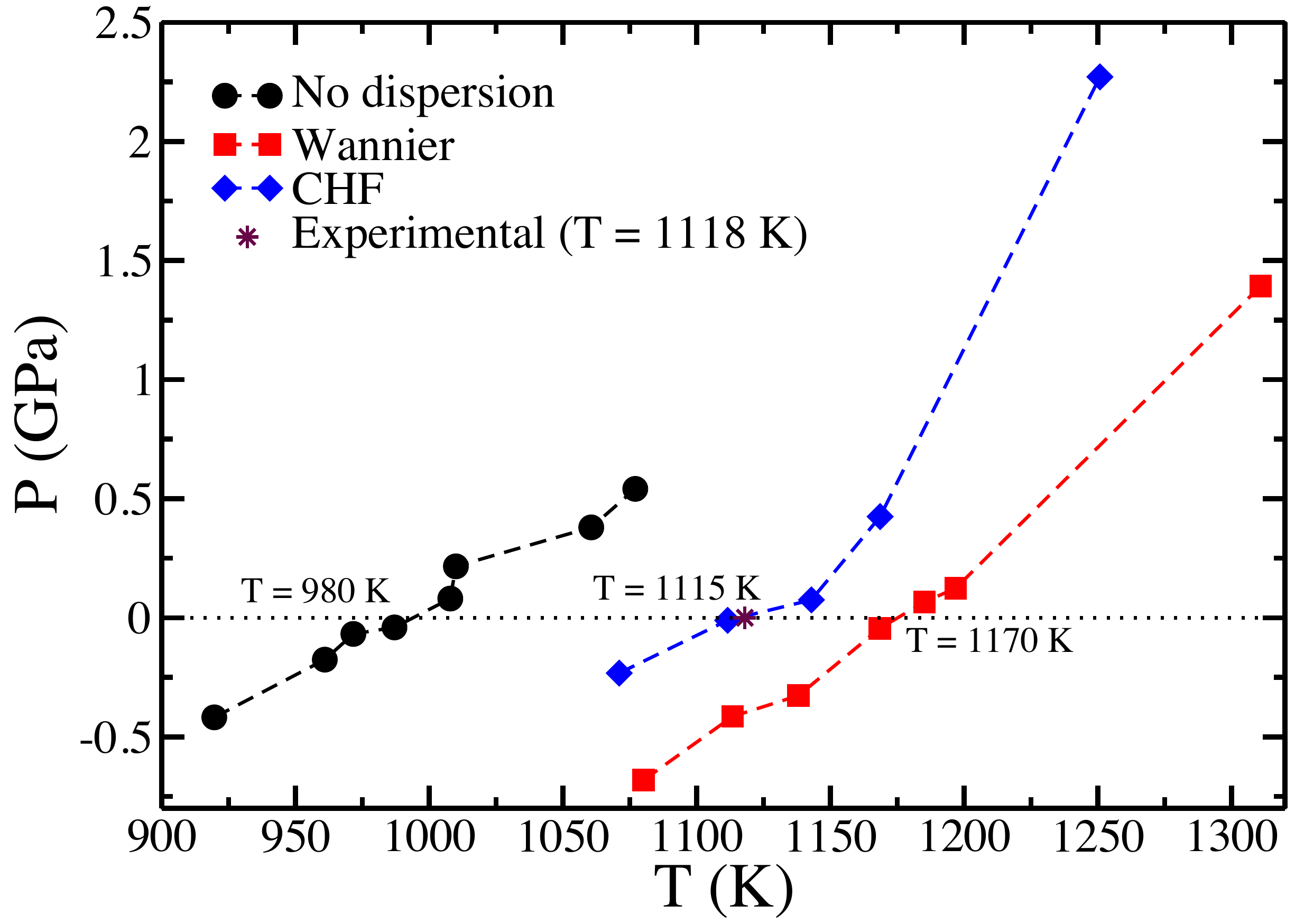}
\end{center}
\caption{\label{fig:meltingpoint} ($P$,$T$) plots for LiF at the solid-liquid coexistence for the three sets of dispersion parameters.}
\end{figure}

The variation of $P$ versus $T$ at coexistence is shown on Figure \ref{fig:meltingpoint} for the three sets of dispersion parameters. At $P=1$~atm, we obtain approximate melting temperatures of 980~K, 1115~K and 1170~K for the case with null dispersion and with the CHF and Wannier parameters, respectively. We then perform long simulations in the $NPT$ ensemble, where the target pressure is set to 1~atm while several target temperatures close to the estimated $T_m$ are studied. 

\begin{figure}[ht!]
\begin{center}
\includegraphics[width=\columnwidth]{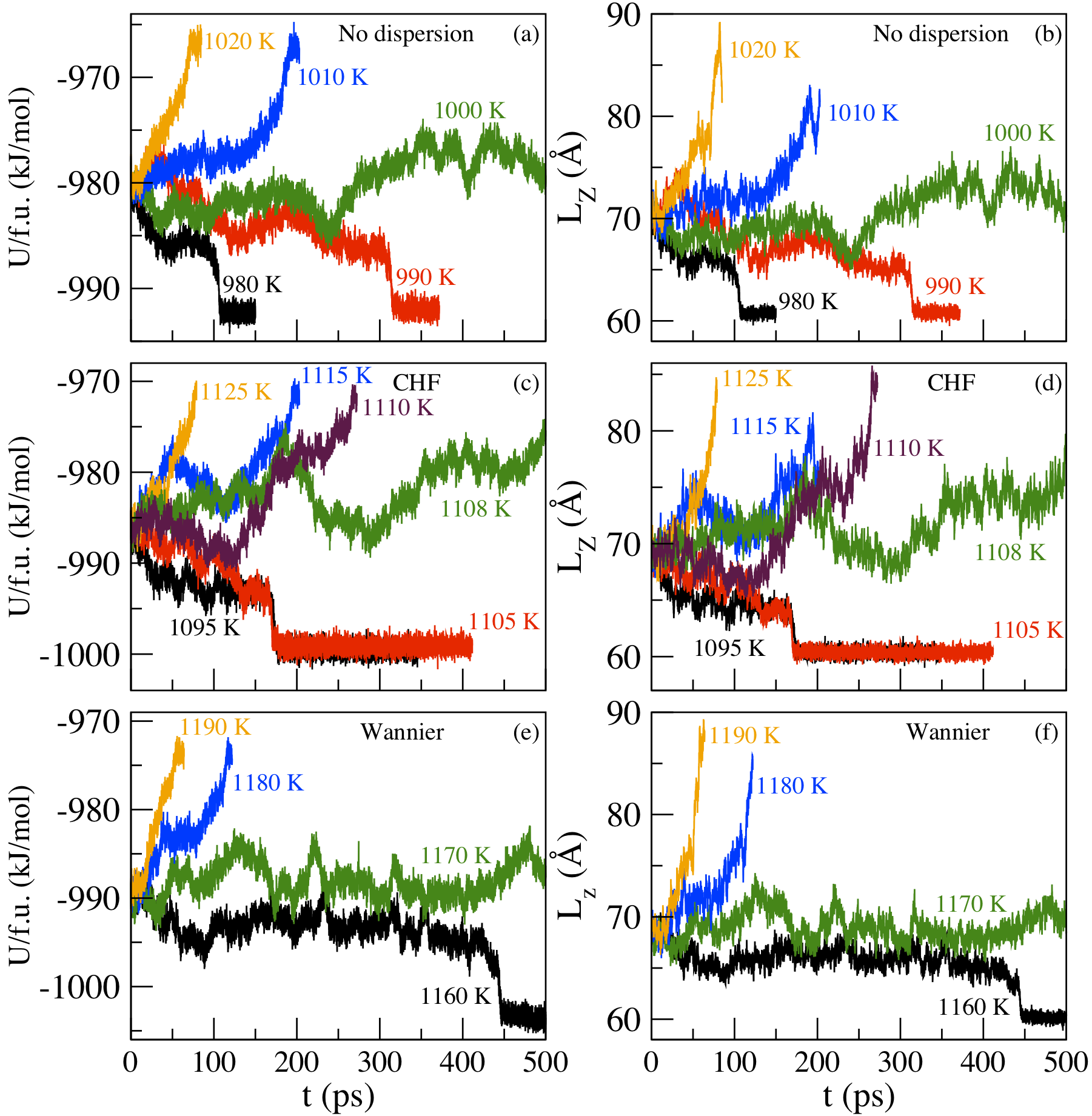}
\end{center}
\caption{\label{fig:meltingpoint2} Variation of the potential energy (left) and cell length along the $z$ axis (right) at several temperatures during $NPT$ simulations performed at several temperatures.}
\end{figure}

We sample the potential energy and the cell length along the $z$ direction during these $NPT$ simulations. The results are shown on Figure \ref{fig:meltingpoint2}. Both quantities remain constant only when the liquid and solid are coexisting~\cite{conde2013a}. We extract refined melting temperatures of 1000~K, 1108~K and 1170~K for the case with null dispersion and with the CHF and Wannier parameters, respectively. This trend is easily understood by the fact that increasing the dispersion effects will result in a stabilization of the most condensed phase, i.e. the solid, with respect to the less condensed phase. Nevertheless, the observed differences are large (up to $>100$~K) despite the fact that no important differences were observed for the single phase structural properties. The CHF case provides a very accurate estimate of the melting point since the experimental value is 1118~K. This result is somewhat surprising since the model using the Wannier dispersion parameters provided the best estimate of the densities of both the solid and liquid phases, again showing the difficulty of building a force field of perfect accuracy for all the properties, even if it includes polarization effects. 

\section{Conclusions}

In conclusion, we have shown in this paper that, although it brings a tiny contribution to the total energy, the dispersion interaction has important effects on the properties of LiF. On the one hand, these effects are not observed in fixed volume simulations of the liquid, for which the structure and the dynamics are independent of the $C_6$ and $C_8$ terms. On the other hand, the predicted equilibrium densities are affected in both the liquid and solid phases: an underestimation of the experimental data by as much as 15~\% is observed when dispersion effects are omitted. But the strongest differences are obtained for the melting point, a quantity which reflects the free energy difference between the solid and liquid phase. This means that as soon as free energy-related quantities have to be evaluated,  special attention must be paid to using a correct parameterization of the dispersion coefficients.

To carry out this parameterization on a purely first-principles basis, which is what we have attempted here, is not straightforward. Here we have tested two sets of parameters. The first one was extracted from condensed-phase Coupled Hartree Fock calculations~\cite{fowler1985b}, while the second one was obtained from the determination of the Maximally Localized Wannier Functions in a simple condensed phase DFT calculation~\cite{rotenberg2010a,silvestrelli2008a}. Both of them provide reasonably good predictions for the densities and the melting points, but more systematic calculations on a series of materials would be needed for identifiying the better parameterization procedure. It is to be expected that the developments which are currently being made for treating accurately dispersion interactions in DFT calculations~\cite{vydrov2010a,vydrov2010b,grimme2004a} will provide useful routes.

\section*{Acknowledgement}
The work has been performed under the HPC-EUROPA2
project (project number: 228398) with the support of the European
Commission -- Capacities Area -- Research Infrastructures.

\section*{References}

\end{document}